\RequirePackage{snapshot}
\documentclass[sigplan,authorversion]{acmart}

\setcopyright{none}
\AtBeginDocument{%
	\providecommand\BibTeX{{%
			Bib\TeX}}}

\acmYear{2025}\copyrightyear{2025}
\setcopyright{acmlicensed}
\acmConference[PPoPP '25]{The 30th ACM SIGPLAN Annual Symposium on Principles and Practice of Parallel Programming}{March 1--5, 2025}{Las Vegas, NV, USA}
\acmBooktitle{The 30th ACM SIGPLAN Annual Symposium on Principles and Practice of Parallel Programming (PPoPP '25), March 1--5, 2025, Las Vegas, NV, USA}
\acmDOI{10.1145/3710848.3710863}
\newcommand{\PHM}[1]{\noindent\textbf{#1}\hspace{0.1em}} % paragraph in the middle
\usepackage[normalem]{ulem}
\usepackage{pifont}
\usepackage{algorithm}
\usepackage{algpseudocode}
\usepackage{graphicx}
\usepackage{float}
\usepackage{subfig}
\usepackage{appendix}
\usepackage{makecell}
\usepackage{bbding}
\usepackage{multirow}
\usepackage{booktabs}
\usepackage{tabularx}
\usepackage{siunitx}
\usepackage{colortbl}
\usepackage{booktabs} 
\usepackage{paralist}
\usepackage{url}
\usepackage{tablefootnote}

\newcommand{\revised}[1]{{#1}}

\definecolor{deepblue}{RGB}{218,215,235}
\definecolor{lightblue}{RGB}{242,241,248}
\usepackage{hyperref}
\usepackage{cleveref}
\crefname{section}{§}{§§}
\Crefname{section}{§}{§§}
\usepackage[normalem]{ulem}
\useunder{\uline}{\ul}{}
\usepackage{listings}
\begin{document}
\title{SGDRC: Software-Defined Dynamic Resource Control for Concurrent DNN Inference on NVIDIA GPUs}
%%%%%%%%%%%%%%%%%%%%%%%%%%%%%%%%%%%%%%%%

%%%%%%%%%%%%%%%%%%%%%%%%%%%%%%%%%%%%%%%%
%%%%%%%% -- PAPER CONTENT STARTS -- %%%%%%%%%

%%
%% end of the preamble, start of the body of the document source.



\author{Yongkang Zhang}
%\authornote{Corresponding author.}
\affiliation{
	\institution{HKUST}
	\city{Hong Kong}
	\country{China}
}

\author{Haoxuan Yu}
\affiliation{
	\institution{HKUST}
	\city{Hong Kong}
	\country{China}
}

\author{Chenxia Han}
\affiliation{
	\institution{CUHK}
	\city{Hong Kong}
	\country{China}
}

\author{Cheng Wang}
\affiliation{
	\institution{Alibaba Group}
	\city{Shenzhen}
	\country{China}
}

\author{Baotong Lu}
\affiliation{
	\institution{Microsoft Research}
	\city{Beijing}
	\country{China}
}

\author{Yunzhe Li}
\affiliation{
	\institution{Shanghai Jiao Tong University}
	\city{Shanghai}
	\country{China}
}

\author{Zhifeng Jiang}
\affiliation{
	\institution{HKUST}
	\city{Hong Kong}
	\country{China}
}

\author{Yang Li}
\affiliation{
	\institution{China University of Geosciences}
	\city{Wuhan}
	\country{China}
}

\author{Xiaowen Chu}
\affiliation{
	\institution{HKUST (Guangzhou)}
	\city{Guangzhou}
	\country{China}
}

\author{Huaicheng Li}
\affiliation{
	\institution{Virginia Tech}
	\city{Blacksburg}
	\country{USA}
}


\renewcommand{\shortauthors}{Zhang et al.}

\begin{abstract}
  Cloud service providers heavily colocate high-priority, latency-sensitive (LS), and low-priority, best-effort (BE) DNN inference services on the same GPU to improve resource utilization in data centers. Among the critical shared GPU resources, there has been very limited analysis on the dynamic allocation of compute units and VRAM bandwidth, mainly for two reasons: (1) The native GPU resource management solutions are either hardware-specific, or unable to dynamically allocate resources to different tenants, or both; (2) NVIDIA doesn't expose interfaces for VRAM bandwidth allocation, and the software stack and VRAM channel architectures are black-box, both of which limit the software-level resource management. These drive prior work to design either conservative sharing policies detrimental to throughput, or static resource partitioning only applicable to a few GPU models.

To bridge this gap, this paper proposes SGDRC, a fully software-defined dynamic VRAM bandwidth and compute unit management solution for concurrent DNN inference services. SGDRC aims at guaranteeing service quality, maximizing the overall throughput, and providing general applicability to NVIDIA GPUs. SGDRC first reveals a general VRAM channel hash mapping architecture of NVIDIA GPUs through comprehensive reverse engineering and eliminates VRAM channel conflicts using software-level cache coloring. SGDRC applies bimodal tensors and tidal SM masking to dynamically allocate VRAM bandwidth and compute units, and guides the allocation of resources based on offline profiling. We evaluate 11 mainstream DNNs with real-world workloads on two NVIDIA GPUs. The results show that compared with the state-of-the-art GPU sharing solutions, SGDRC achieves the highest SLO attainment rates (99.0\% on average), and improves overall throughput by up to 1.47$\times$ and BE job throughput by up to 2.36$\times$.
\end{abstract}

%\begin{CCSXML}
%\begin{CCSXML}
%	<ccs2012>
%	<concept>
%	<concept_id>10002951.10003227.10010926</concept_id>
%	<concept_desc>Information systems~Computing platforms</concept_desc>
%	<concept_significance>500</concept_significance>
%	</concept>
%	<concept>
%	<concept_id>10010520.10010521.10010537.10003100</concept_id>
%	<concept_desc>Computer systems organization~Cloud computing</concept_desc>
%	<concept_significance>500</concept_significance>
%	</concept>
%	</ccs2012>
%\end{CCSXML}

%\ccsdesc[500]{Information systems~Computing platforms}
%\ccsdesc[500]{Computer systems organization~Cloud computing}



\maketitle

%-------------------------------------------------------------------------------

\section{Introduction}\label{sec:introduction}

With rapid technological advancements in machine intelligence across various fields such as vision recognition~\cite{He2016,Huang2017} and natural language processing~\cite{Vaswani2017,Devlin2019,OpenAI2023}, an increasing number of industries are deploying large-scale deep neural network (DNN) inference services in cloud data centers to support their businesses. The first-class citizens are latency-sensitive (LS) services, which have high priority and stringent requirements on tail latency (e.g., autonomous driving~\cite{Hu2023}). In contrast, other services running in the cloud are typically throughput-oriented batch jobs served in a best-effort (BE) manner due to their low priority, such as artificial intelligence generative tasks~\cite{Esser2024}.

To enhance GPU utilization, it is a common practice to share GPUs among multiple LS tasks~\cite{Gujarati2020,Romero2021,Choi2022,Li2023,Zhang2023a} or between LS and BE tasks~\cite{Xiao2020,Zhang2022c,Weng2022,Zhang2022a} using NVIDIA's native support~\cite{nvidia-mps, nvidia-mig}.
One effective method is to employ the Multi-Process Service (MPS)~\cite{nvidia-mps}, which enables the concurrent execution of GPU kernels from different tasks on the same GPU instance.
Additionally, GPUs can be partitioned into several distinct logical instances with guaranteed and isolated resources using Multi-Instance GPU (MIG)~\cite{nvidia-mig}. 

Unfortunately, neither MPS nor MIG is perfect. For example, MPS statically partitions GPUs at the thread slice level and cannot isolate VRAM bandwidth, resulting in unmanaged contention among colocated services. While MIG fully isolates compute units and VRAM bandwidth, it is available only in a few flagship GPUs (e.g., A100 and H100) but not in other low-end GPUs (e.g., Tesla T4), which many IT giants use to deploy DNNs in order to reduce the total cost of ownership (TCO)~\cite{Yu2022}. Besides, its granularity is too coarse (e.g., up to 7 instances of 10 GiB for A100) and can only reconfigure the allocation when it is idle. This raises the question of whether we can develop \emph{a widely applicable GPU sharing scheme that provides strong performance isolation and dynamic allocation of resources}. Indeed, several studies have explored this issue, as summarized in Fig.~\ref{fig:illustration_multiplexing}a$\sim$d:

\begin{figure*}[t]
	\centering
	\includegraphics[width=\textwidth]{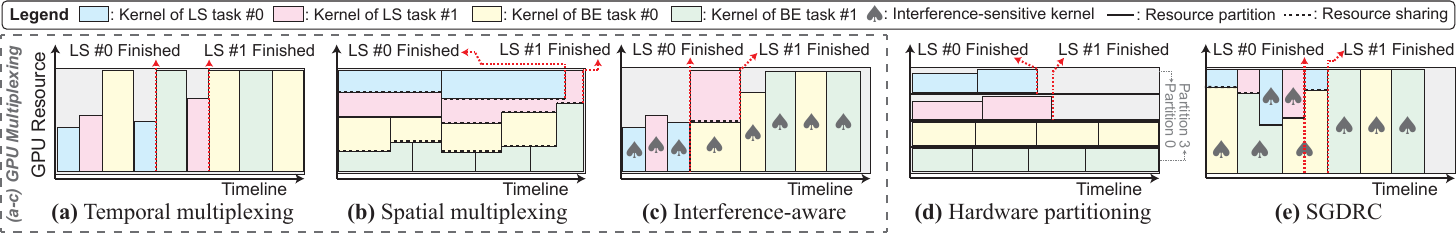}
	\vspace{-0.3in}
	\caption{Illustration of existing GPU sharing schemes and SGDRC. The gray (or colored) rectangles represent GPU resources (or DNN kernels). The width (or height) of a colored rectangle represents the runtime (or resource utilization) of a DNN kernel.}
	\label{fig:illustration_multiplexing}
\end{figure*} 

a) \textit{Temporal multiplexing} (e.g., TGS~\cite{Wu2023}) allows only one DNN to be executed exclusively on the GPU at a time~\cite{Capodieci2018,Gujarati2020,Wu2023,Gujarati2020} to ensure low latency for LS tasks. 

b) \textit{Spatial multiplexing} (e.g., Reef~\cite{Han2022}) uses MPS~\cite{Cui2021} or kernel padding~\cite{Wu2015,Jain2019,Han2022,Zhao2023} to enable concurrent execution of multiple DNNs on a GPU and maximize throughputs.

c) \textit{Interference-aware multiplexing} (e.g., Orion~\cite{Foteini2024}) is based on spatial multiplexing, which predicts the interference among colocated tasks and only allows the coexecution of mildly interfering kernels. This ensures the low latency of LS services while achieving higher throughput for BE tasks compared to temporal multiplexing. 

d) \textit{Software-controlled hardware partitioning} (i.e., Fractional GPU, FGPU~\cite{Jain2019}) eliminates inter-task resource interference by statically partitioning the GPU's compute units and VRAM channels at the software level (\S\ref{subsec:hardware_patitioning_limitations}).

Our characterization with realistic workloads (\S\ref{subsec:limitation_existing_gpu_sharing}) indicates that \textit{these approaches cannot achieve both low latency for LS services and high throughput for BE tasks when they are colocated}. Although hardware partitioning (FGPU~\cite{Jain2019}) isolates SMs and VRAM channels, \textit{it remains impractical for most GPUs (including newer models)} due to its strong assumptions about the GPU's VRAM channel mapping function (which does not hold for most GPUs) and sensitivity to noise introduced by black-box GPU cache policies (\S\ref{subsec:hardware_patitioning_limitations}). Additionally, its cache coloring is not suitable for newer GPU architectures (\S\ref{subsec:findings-vram-channel}).

To flip the status quo, we propose SGDRC\footnote{SGDRC stands for \underline{S}oftware-defined \underline{G}PU \underline{D}ynamic \underline{R}esource \underline{C}ontrol, the GPU equivalent of Intel CPU's Dynamic Resource Control (DRC)~\cite{Zhang2021}.}, a novel GPU sharing solution for concurrent DNN inference generally applicable to NVIDIA GPUs (Fig.~\ref{fig:illustration_multiplexing}e). Our key intuition is to use software-defined dynamic hardware partitioning to eliminate scheduling constraints of interference-aware multiplexing and achieve both responsiveness for LS services and high overall throughput. To this end, we need to address three fundamental challenges: 

1) NVIDIA's GPU architecture is opaque, and its VRAM channel mapping is still publicly unknown.

2) NVIDIA doesn't provide any interfaces to control VRAM bandwidth assignment, and its proprietary library and driver disallow partitioning VRAM bandwidth among different tasks.

3) NVIDIA GPU's VRAM channel mapping is frozen in the hardware, making it difficult to dynamically allocate VRAM bandwidth to tasks during the GPU's runtime.

SGDRC resolves these challenges by:

1) Conducting, to the best of our knowledge, the first full-spectrum reverse engineering, \textit{unearthing the general VRAM channel structure of black-box NVIDIA GPUs} (\S\ref{subsec:label-vram-channels}$\sim$\ref{subsec:findings-vram-channel}).

2) Leveraging DNNs to \textit{learn VRAM channel hash mapping} without any assumptions on hash function structure, which is tolerant of the GPU cache noise (\S\ref{subsec:fit-hash-mapping}).

3) Reducing inter-task VRAM channel conflicts with \textit{low-overhead, fine-grained page coloring at the software level}, which is generally applicable to NVIDIA GPUs (\S\ref{sec:vram-bandwidth-allocation}).

4) \textit{Dynamically allocating VRAM bandwidth and compute units} during runtime (\S\ref{sec:elastic-sm-allocation}).

Our experiments demonstrate that, compared with state-of-the-art GPU sharing solutions, SGDRC achieves the highest SLO attainment rates (99.0\% on average), and improves overall throughput by up to 1.47$\times$ and BE job throughput by up to 2.36$\times$.

\section{Background}\label{sec:background}

\subsection{A Primer on NVIDIA GPUs and Software Stack}\label{subsec:nvidia-gpu-arch}

\begin{figure}
	\centering
	\includegraphics[width=1\columnwidth]{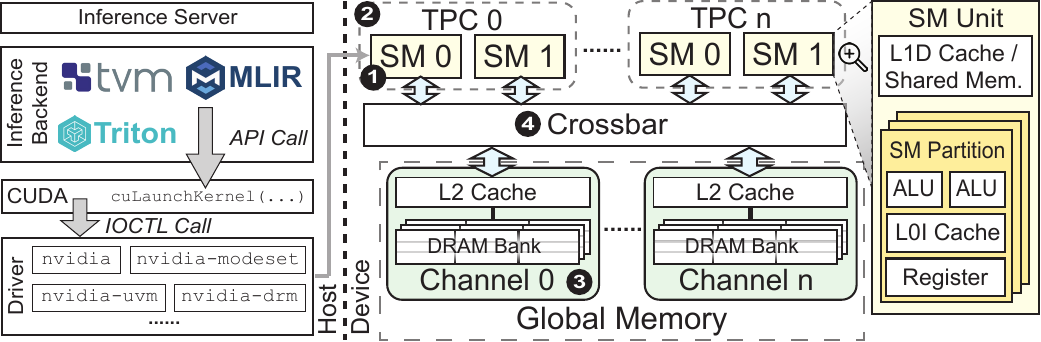}
	\caption{NVIDIA GPU's architecture and the software stack.}
	\label{fig:gpu_memory_arch}
	\vspace{-.1in}
\end{figure}

Based on NVIDIA's official documents~\cite{gp100-fbpa,gp100-mmu-format} and previous work on reverse-engineering GPUs~\cite{Amert2017,Mei2017,Jain2019,Zhang2023,envytools}, Fig.~\ref{fig:gpu_memory_arch} illustrates NVIDIA's GPU architecture and software stack:

\PHM{Software stack.} User-space programs interact with NVIDIA GPUs by calling APIs (e.g., \texttt{cuLaunchKernel}) from their closed-source libraries (e.g., CUDA). These libraries forward the requests to NVIDIA's driver modules. Most modules are closed-source, while a few are open-source, including \texttt{nvidia-uvm}, a module that manages the GPU's unified memory.

\PHM{Compute units.} A \textit{Streaming Multiprocessor} (\textit{SM}, \ding{182}) is the basic compute unit, containing multiple \textit{SM partitions} (\textit{SMP}). A \textit{Texture Processing Cluster} (\textit{TPC}, \ding{183}) contains two \textit{SMs}. \textit{Threads} from \textit{kernels} (GPU functions) are scheduled to different \textit{SMPs}.

\PHM{Memory hierarchy.} NVIDIA GPUs' memory is divided into three levels~\cite{Mei2017}: 1) L1 data cache and shared memory (private to each SM); 2) L2 unified cache (shared by all SMs); and 3) video RAM (VRAM, shared by all SMs). VRAM is composed of multiple GDDR units. Each unit has a set of Miss Status Holding Registers (MSHRs) and multiple DRAM banks and maps to a set of L2 cache (referred to as a \textit{VRAM channel}~\cite{gp100-fbpa}, \ding{184}). Within the \textit{Crossbar} (\ding{185}), there is a direct bus between each SM and each L2 cache controller. This implies that the latency for any SM to access the L2 cache on any channel is the same, making NVIDIA GPU a \revised{Uniform} Memory Access (UMA) architecture. Each physical address is mapped to a VRAM channel, an L2 cacheline, and a DRAM bank row through black-box hash mapping functions implemented in gate circuits~\cite{Jain2019}. These functions ensure that the physical VRAM addresses are evenly mapped to each VRAM channel and thus maximize the VRAM throughput when kernels read from / write to the global memory.

\subsection{Resource Contention in GPU Sharing}\label{subsec:characterize-conflicts}

Colocating multiple DNN workloads on the same GPU leads to contention for the following resources:

\begin{figure}[tb]
	\centering
	%\vspace{-.15in}
	
	\includegraphics[width=1.0\columnwidth]{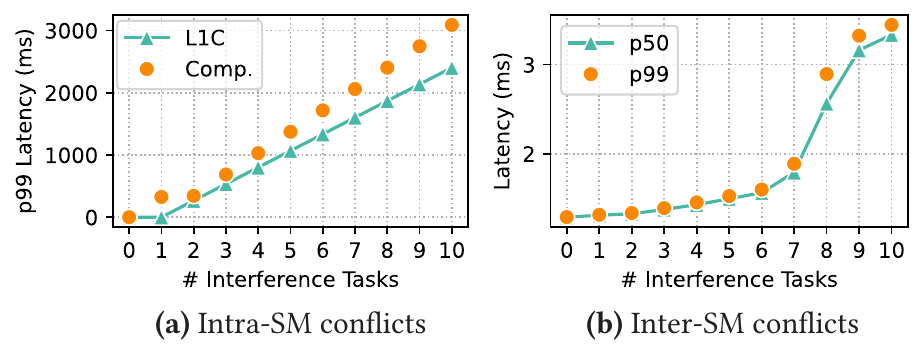}

	% \vspace{-.15in}
	\caption{Resource contention in GPU sharing. We measure the p99 latency of the victim task to quantify the interference. \textit{L1C} (or \textit{Comp.}) in \textbf{(a)} denotes the introduction of \textit{L1 cache} (or \textit{compute unit}) interference. \textbf{Testbed:} RTX A2000.}
\vspace{-.13in}
\label{fig:explore-gpu-sharing-conflicts}
\end{figure}

\PHM{Intra-SM conflicts.} Kernel block threads from different tenants running on the same SM could contend for intra-SM resources. For instance, when all Floating-Point Units (FPUs) are actively processing, additional floating-point operations experience delays, hindering the progress of other kernels. In addition to computational units, warps (groups of threads) on the same SM also compete for SM-local memory resources, such as the L1 cache, shared memory, and instruction cache. We colocate one victim task with multiple interference tasks and concurrently execute them on the RTX A2000 to quantify the interference. The victim task and the compute unit interference tasks perform matrix multiplications. The L1 cache interference tasks repeatedly populate the L1 data cache. All tasks share SMs to contend for intra-SM resources. The results are presented in Fig.~\ref{fig:explore-gpu-sharing-conflicts}(a).

\PHM{Inter-SM contention.} As described in \S\ref{subsec:nvidia-gpu-arch}, different SMs share all VRAM channels. Consequently, physical addresses accessed by threads in different SMs may map to the same VRAM channel. Simultaneous access to these addresses leads to frequent contention for the limited L2 cache space and MSHRs. Additionally, since a DRAM bank can only serve one request in a clock cycle, memory requests from multiple threads to the same DRAM bank must be processed sequentially~\cite{Jain2019}, increasing VRAM access latency. To demonstrate these conflicts, we concurrently execute the victim task and interference tasks on the RTX A2000. The victim task performs matrix multiplications, while interference tasks continuously read from and write to VRAM addresses to create L2 cache misses. SMs are divided into different tasks to eliminate intra-SM contention using NVIDIA MPS~\cite{nvidia-mps}. The results are presented in Fig.~\ref{fig:explore-gpu-sharing-conflicts}b.

\section{Motivation and Related Work}\label{sec:motivation}

\subsection{Limitations of GPU Multiplexing Solutions}\label{subsec:limitation_existing_gpu_sharing}

\begin{figure}[tb]
	\centering
	\vspace{-.15in}
	\subfloat[\revised{Temporal multiplexing}]
	{
		\includegraphics[width=0.49\columnwidth]{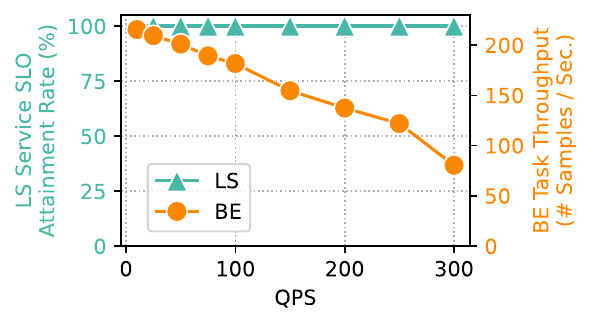}
		\label{fig:temporal_evaluation}
	}
	\subfloat[\revised{Spatial multiplexing}]
	{
		\includegraphics[width=0.49\columnwidth]{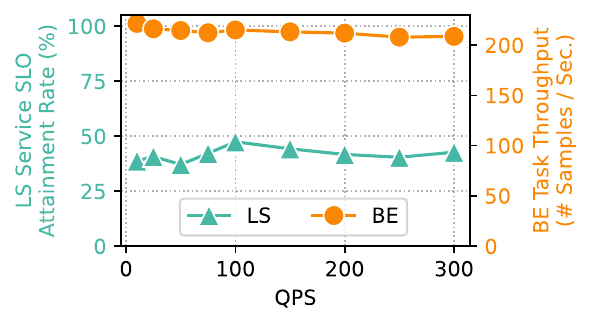}
		\label{fig:spatial_evaluation}
	}
	% \vspace{-.15in}
	\caption{Limitations of GPU temporal and spatial multiplexing. \textbf{(a)} Temporal multiplexing~\cite{Wu2023,Gujarati2020} cannot achieve high throughput for BE tasks; \textbf{(b)} Spatial multiplexing~\cite{Zhao2023} can achieve high throughput, but at the cost of sacrificing the LS task's SLO attainment rate (defined in \S\ref{subsec:final_experimental_setup}) due to resource contention; \textbf{LS workload:} MobileNet V3; \textbf{BE workload:} DenseNet161; \textbf{Testbed: } RTX A2000.}
	\vspace{-.1in}
	\label{fig:todo}
\end{figure}

\begin{figure}
	\centering
	\includegraphics[width=1\linewidth]{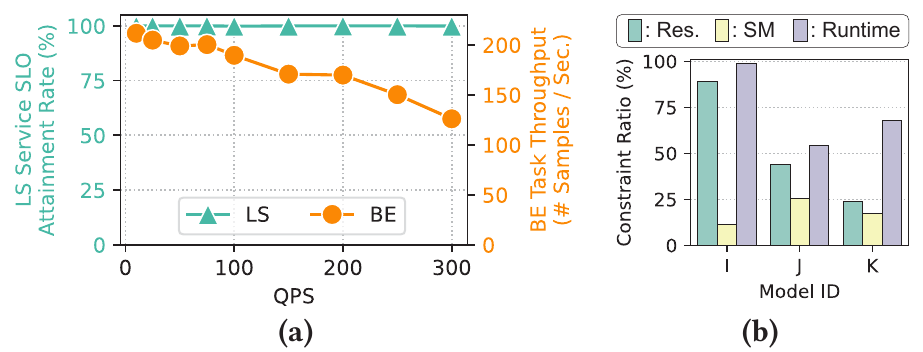}
	\vspace{-0.3in}
	\caption{\revised{Interference-aware multiplexing} is not panacea. \textbf{(a)} As the load increases, the LS service maintains high SLO attainment rate. However, the throughput of BE task substantially declines. \textbf{LS Workload:} MobileNet V3; \textbf{BE Workload:} DenseNet161; \textbf{Testbed: } RTX A2000. \textbf{(b)} Analysis of scheduling constraints of BE tasks (I $\sim$ K in Tab.~\ref{tab:list_of_testing_models}, running on RTX A2000). \textbf{Res.:} Constraints on SM or VRAM bandwidth utilization; \textbf{SM:} Constraints on the required number of SMs; \textbf{Runtime:} Constraints on kernel runtime.}
	\label{fig:orion-example}
\end{figure}

\textit{Temporal multiplexing} eliminates contention in shared GPU resources and meets the low-latency requirements of LS services~\cite{Capodieci2018,Zhou2018}, but cannot fully harness the GPU's resources, as BE tasks may be starved due to frequent LS task preemption, leading to undesirable throughput~\cite{Han2022} (Fig.~\ref{fig:temporal_evaluation}).

\textit{Spatial multiplexing} leads to intra- and inter-SM conflicts when co-executing LS and BE kernels (Fig.~\ref{fig:spatial_evaluation})~\cite{Yandrofski2022}.

\revised{Compared to temporal multiplexing and spatial multiplexing, \textit{interference-aware multiplexing} (e.g., Orion~\cite{Foteini2024}) achieves a better trade-off between low latency of LS services and high throughput.} However, it is not perfect in all scenarios. We take Orion~\cite{Foteini2024} as an example. As the load of the LS service increases, the large number of executing LS kernels poses challenges for Orion's scheduler in selecting suitable BE kernels for co-execution. As a result, despite the LS service maintaining a high SLO attainment rate, the throughput of BE tasks decreases as the concurrency of LS services increases (Fig.~\ref{fig:orion-example}a).

This is because Orion imposes numerous constraints on the co-execution of BE kernels to ensure low latency for LS tasks, as it cannot avoid intra-SM and inter-SM interference. These constraints limit the BE task throughput. For the BE models listed in Tab.~\ref{tab:list_of_testing_models}, 73.8\% of their kernels are subjected to at least one constraint (Fig.~\ref{fig:orion-example}b). These excessive constraints lead to decreased BE task throughput as the load of the LS service increases. Furthermore, relaxing any constraint increases the LS task's latency, which means that they are all indispensable for maintaining LS service responsiveness.

\subsection{Limitations of GPU Partitioning}\label{subsec:hardware_patitioning_limitations}

Although existing solutions all tried to work around GPU resource partitioning, they result in either undesirable DNN inference performance, low GPU utilization, or both (\S\ref{subsec:limitation_existing_gpu_sharing}). Unfortunately, software-controlled GPU hardware partitioning is also flawed. While there are numerous mature software~\cite{Jain2019,Han2022,Ng2023} or hardware-based~\cite{Otterness2020,Bakita2023,Chow2023} solutions for partitioning computational units, VRAM channel allocation remains challenging due to its close coupling with proprietary GPU hardware and driver implementations.

Fractional GPU (FGPU)~\cite{Jain2019} stands out as the only software-based GPU sharing solution capable of statically partitioning both compute units and VRAM channels on GTX 1080. It isolates VRAM channels using cache coloring, a technique employed in CPU last-level cache isolation.

\PHM{FGPU is inapplicable to most commodity GPUs and new GPU architecture.} FGPU assumes that the GPU L2 cacheline and DRAM bank hash mapping functions are pure XOR functions. We attempted to reverse engineer other GPUs using FGPU's approach, but all failed because this assumption does not hold for many NVIDIA GPUs. The key issue lies in the nature of the XOR function: it is \textit{linear}, mapping a VRAM space of size $2^N$ bytes to $2^M$ VRAM channels. However, many GPUs use \textit{non-linear} VRAM channel hash mappings, which map a VRAM space of arbitrary size to an arbitrary number of VRAM channels. This non-linearity arises because the VRAM size and the number of VRAM channels are often not powers of 2 (Tab.~\ref{tab:num_vram_channels_each_gpu}). The number of VRAM channels can be cross-validated by the number of GDDR chips on the GPU (Fig.\ref{fig:gpu-disassembly-photo} in \S\ref{appendix:reverse-engineering-findings}) and the theoretical calculation (i.e., VRAM bus width divided by the bus width per memory unit). Furthermore, FGPU only supports page coloring based on 4 KiB granularity, the minimum page size supported by NVIDIA GPU's Memory Management Unit (MMU)~\cite{gp100-mmu-format}. However, this is inapplicable to newer GPU architectures (\S\ref{subsec:findings-vram-channel}). Moreover, FGPU's reverse engineering approach is not tolerant to cache noise. Even one false positive sample can pollute the equation system and the reverse-engineered hash function. However, we find that in Pascal and Ampere GPUs, around 1\% and 5\% sampled conflicted addresses are false positives due to the black-box GPU cache policies.

\begin{table}[]
	\caption{VRAM size, VRAM bus width, and \# VRAM channels of 3 GPUs. FGPU~\cite{Jain2019} is only compatible with \textit{GTX 1080}.}
	\scriptsize
	\centering
	\begin{tabular}{c|ccccc}
		\hline
		\textbf{Specifications} & \textit{\textbf{GTX 1080}}  & \textbf{Tesla P40} & \textbf{RTX A2000}   \\ \hline
		Architecture                             & Pascal               & Pascal          & Ampere         \\
		VRAM size (GiB)                         & 8                      & 24             & 12              \\
		%L2 Cache Size (MiB)                         & 2                      & 6                      & 3             & 3              & 6             \\
		VRAM bus width (bit)                         & 256                      & 384             & 192               \\
		\makecell[c]{Bus width per GDDR unit (bit)}                         & 32                      & 32             & 32            \\
		\# VRAM channels                         & 8                      & 12             & 6               \\ \hline
	\end{tabular}
\label{tab:num_vram_channels_each_gpu}
\vspace{-.1in}
\end{table}

\PHM{FGPU cannot scale GPU resource allocation.} When LS tasks are inactive, FGPU cannot utilize idle compute units, which adversely affects the throughput of BE tasks. Additionally, statically allocating a portion of VRAM channels to tasks is not always the optimal choice, as the reduced and fixed L2 cache size leads to an increase in L2 cache misses (as verified in FGPU's experiments~\cite{Jain2019}).

%\subsection{State-of-the-Arts of GPU Sharing}\label{subsec:motivation_takeaway}

\begin{table*}[t]  % 使用table*环境创建跨越两栏的表格，选项[t]表示表格在页面的顶部
	\centering  % 使表格内文本居中
	\caption{A comparison of mainstream GPU sharing solutions. \textbf{Note:} MPS~\cite{nvidia-mps} partitions compute units at the thread level.}%\textit{Temp.+Spat.} and \textit{Inter.-aware} refers to temporal-spatial multiplexing and interference-aware scheduling, respectively.}
% \footnote{\revised{As MPS and MIG are close-sourced, their PCIe isolation capability is unknown.}
	\scriptsize
	
	\begin{tabular}{c|ccccccccc}
		\hline \textbf{Method}
		& \makecell[c]{\textbf{GPU Sharing}\\\textbf{Scheme}} & \makecell[c]{\textbf{Implementation}} & \makecell[c]{\textbf{Support All}\\\textbf{NVIDIA}\\\textbf{GPUs}} &  \makecell[c]{\textbf{Computing}\\\textbf{Unit}\\\textbf{Partitioning}} & \makecell[c]{\textbf{VRAM}\\\textbf{Bandwidth}\\\textbf{Partitioning}} & \makecell[c]{\textbf{Compute Unit}\\\textbf{Dynamic}\\\textbf{Allocation}} & \makecell[c]{\textbf{VRAM B.W.}\\\textbf{Dynamic}\\\textbf{Allocation}}  & \makecell[c]{\textbf{Reconfiguration}\\\textbf{Overhead}} \\
		\hline
		MPS~\cite{nvidia-mps}    & Native       & Hardware   & \Checkmark & \Checkmark & \XSolidBrush   & \XSolidBrush  & \XSolidBrush & High \\
		MIG~\cite{nvidia-mig}    & Native       & Hardware    & \XSolidBrush & \Checkmark & \Checkmark  & \XSolidBrush  & \XSolidBrush & High \\
		FGPU~\cite{Jain2019}     & Hardware partitioning       & Driver  & \XSolidBrush & \Checkmark & \Checkmark & \XSolidBrush  & \XSolidBrush  & High \\
		TGS~\cite{Wu2023}        & Temporal multiplexing      & User-space & \Checkmark & \revised{N/A} & \revised{N/A}   & \Checkmark  & \XSolidBrush  & Low \\
		%Abacus~\cite{Cui2021}    & Spatial & User-space & Strong & \XSolidBrush    & \Checkmark   & \Checkmark    & \XSolidBrush  \\
		%iGniter~\cite{Xu2023}    & Spatial & User-space & Medium & \XSolidBrush    & \Checkmark   & \Checkmark    & \XSolidBrush  \\
		Reef~\cite{Han2022}      & Spatial multiplexing       & Driver   & \XSolidBrush  & \Checkmark   & \XSolidBrush   & \Checkmark   & \XSolidBrush  & Medium \\
		\revised{Paella}~\cite{Ng2023} & \revised{Spatial multiplexing}   & User-space & \revised{\Checkmark}  & \revised{\Checkmark} & \XSolidBrush  & \Checkmark   & \XSolidBrush  & Medium \\
		%KRISP~\cite{Chow2023}      & Spatial       & User-space        & \Checkmark  & \XSolidBrush & \XSolidBrush    & \Checkmark   & \Checkmark    & Fine  \\
		Orion~\cite{Foteini2024} & Interference-aware   & User-space & \revised{\Checkmark} & \XSolidBrush & \XSolidBrush & \XSolidBrush  & \XSolidBrush  & Low \\
		KRISP~\cite{Chow2023} & Spatial multiplexing & Driver & \XSolidBrush & \Checkmark  & \XSolidBrush  & \Checkmark   & \XSolidBrush  & Low \\
		SGDRC (\textbf{Ours})                     & Dynamic partitioning       & User. + Driver    & \Checkmark & \Checkmark & \Checkmark     & \Checkmark   & \Checkmark   & Low \\
		\hline
	\end{tabular}
	\label{tab:dnn_sharing}
\end{table*}

\subsection{Related Work}

\S\ref{subsec:limitation_existing_gpu_sharing} and \ref{subsec:hardware_patitioning_limitations} emphasize the bottleneck of GPU sharing. Here, we summarize the most competitive \revised{GPU sharing solutions for DNN inference} in Tab.~\ref{tab:dnn_sharing}. Differing from these works, SGDRC is \textit{the only solution simultaneously achieving the following goals}:

\PHM{Dynamic GPU compute unit partitioning.} \revised{GPU compute unit partitioning has been supported at both the hardware and software levels. Both NVIDIA~\cite{tmd-patent,Bakita2023} and AMD~\cite{Otterness2020} have exposed their hardware interfaces to control a kernel's SM placement.} SGDRC is the first work that \textit{leverages NVIDIA's little-known official interface~\cite{tmd-patent} to enable dynamic GPU compute unit allocation}. Reef~\cite{Han2022} and Paella~\cite{Ng2023} use kernel padding, a widely used software-based SM partitioning approach, to partition SM units. This technique merges multiple kernels into one monolithic kernel for co-execution, which severely limits concurrency because it requires the colocated BE kernel's runtime to be smaller than the LS kernel's. It can help SGDRC extend to other vendors' GPUs without official SM masking interface support.

\PHM{GPU VRAM channel reverse engineering.} Many works have reverse-engineered the hash mapping function of CPU last-level cache (LLC)~\cite{Apecechea2015,Maurice2015,Farshin2019} and partitioned LLC for different tenants using cache coloring~\cite{Scolari2016,Ye2014,Farshin2019}, which is inapplicable to GPUs, because they leveraged the CPUs' non-uniform cache access latency feature, while GPUs are \textit{UMA models}. FGPU~\cite{Jain2019} cracks the GPU's VRAM channel mapping function by assuming that this function is a pure XOR function, which is inapplicable to most GPU models (\S\ref{subsec:hardware_patitioning_limitations}). SGDRC \textit{first reveals a general VRAM channel mapping structure of NVIDIA GPUs} without relying on the NUMA (Non-uniform Memory Access) feature to mark the channel IDs or assumptions about the hash function structure. 

\PHM{Dynamic GPU VRAM bandwidth allocation.} Although an existing work~\cite{Zhao2023a} proposed a new GPU architecture in the simulator to support VRAM bandwidth allocation, GPU VRAM bandwidth allocation is much more challenging than compute unit partitioning on commodity GPUs, because only NVIDIA has so far implemented static partitioning (i.e., MIG~\cite{Li2022}) for VRAM channels on a few flagship server GPUs (e.g., A100, H100). FGPU~\cite{Jain2019} only implemented static VRAM channel isolation on the GTX 1080 and is inapplicable to new GPU architectures (\S\ref{subsec:hardware_patitioning_limitations}). Unlike these approaches, SGDRC enables dynamic VRAM channel allocation for all NVIDIA GPUs \textit{without hardware modifications} and can be \textit{easily reconfigured} by simply moving tensors to map them to other VRAM channels.

It is essential to clarify that MIG~\cite{nvidia-mig} complements our work, as SGDRC provides \textit{dynamic resource allocation for low-end GPUs}. \revised{Both Paella~\cite{Ng2023} and Reef~\cite{Han2022} are orthogonal to our work.} Reef~\cite{Han2022} primarily focuses on achieving fast BE task preemption. \revised{The primary contribution of Paella~\cite{Ng2023} lies in optimizing low-latency GPU kernel scheduling for DNN inference based on spatial multiplexing.} KRISP~\cite{Chow2023} is an elastic compute unit allocator tailored for AMD GPUs based on AMD's open-source GPU driver. In contrast, SGDRC focuses on \textit{dynamic VRAM channel and compute unit allocation} for DNN inference tasks on \textit{NVIDIA GPUs}.

%\subsection{Key Observations}

%After elaborating drawbacks of state-of-the-arts, now we present the key insights that drive our research.

%\PHM{Bursty and unbalanced GPU utilization.} Throughout the execution of a DNN inference task, both compute unit and VRAM bandwidth utilization display bursty patterns. LS tasks notably underutilize the GPU (Fig. \ref{fig:memcompintensitytimeline}) \cite{Foteini2024}. The bursty nature of GPU utilization and the significant underutilization of the GPU emphasize the importance of sharing the GPU among multiple LS and BE DNN inference tasks.

%\begin{figure}
%	\centering
%	\includegraphics[width=1\linewidth]{sections/motivation/figures/mem_comp_intensity_timeline.pdf}
%	\caption{GPU resource utilization of MobileNetV3 (LS task) and DenseNet161 (BE task) on Tesla V100.}
%	\label{fig:memcompintensitytimeline}
%	\vspace{-.1in}
%\end{figure}

%\PHM{Interference-sensitive kernels limit task colocation.} \todo{TODO.}

%\PHM{VRAM bandwidth dynamic allocation is feasible.} \todo{TODO.}

\section{SGDRC Overview}\label{sec:missile-overview}

Based on the analysis of four GPU sharing approaches, SGDRC employs a completely different method compared to previous works: enabling dynamic resource partitioning at the software level to eliminate constraints for kernel colocation. We design SGDRC based on the following principles:

1) \textit{Fully software-defined}, with no hardware modification.

2) \textit{High elasticity} to maximize overall throughput.

3) \textit{Generally applicable to NVIDIA GPUs.}

\begin{figure}
	\centering
	\includegraphics[width=1.0\columnwidth]{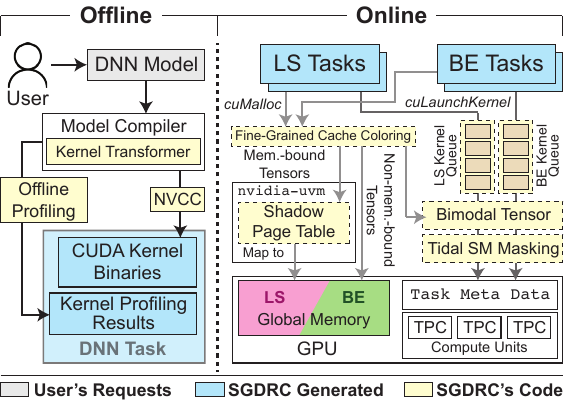}
	\vspace{-0.1in}
	\caption{SGDRC's architecture. Modules with dashed borders are on the critical path of DNN inference.} %\textit{Italic texts} represent intercepted CUDA API calls.}
	\label{fig:missile-arch}
\end{figure}

\begin{figure}
	\centering
	\includegraphics[width=1\linewidth]{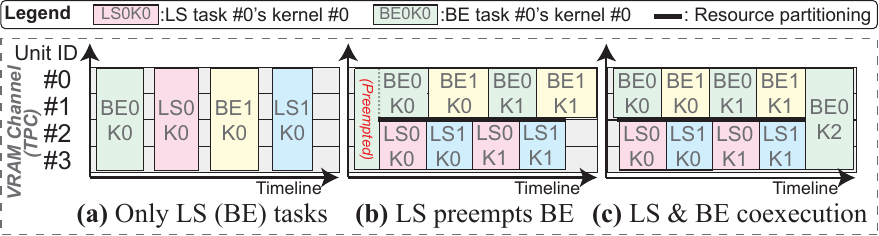}
	\vspace{-0.2in}
	\caption{SGDRC's dynamic VRAM channel and compute unit allocation. The example has 4 TPCs (or VRAM channels).} %$LS_iK_j$ ($BE_iK_j$) represents the $j$-th kernel of the $i$-th LS (BE) task.}
	\label{fig:spatial-temporal-multiplexing-illustrate}
\end{figure}

SGDRC serves user-submitted DNN models in two phases (Fig.~\ref{fig:missile-arch}). 1) \textit{Offline phase}: Users submit their models (e.g., in ONNX or PyTorch format) to SGDRC. SGDRC leverages DNN compilers (e.g., TVM~\cite{Chen2018}, MLIR~\cite{Lattner2021}, and Triton~\cite{Tillet2019}) to fuse and compile DNN operators, then transforms the CUDA kernels to enable VRAM channel dynamic allocation, and uses \texttt{nvcc} to generate \texttt{cubin} binaries. It offline profiles kernels' VRAM bandwidth consumption and the minimum number of required SMs to make dynamic allocation decisions. 2) \textit{Online phase}: SGDRC eliminates resource conflicts (\S\ref{subsec:characterize-conflicts}) in the following ways:

i) \textbf{Inter-SM conflicts.} Through extensive reverse engineering of the black-box GPU architecture, SGDRC finds a general way to crack the VRAM channel mapping (\S\ref{sec:vram-channel-reverse-engineering}). When LS and BE tasks are co-executed, SGDRC allocates $1-Ch_{BE}$ (or $Ch_{BE}$) of VRAM channels to memory-bound LS (or BE) tensors using \textit{shadow page table} and \textit{bimodal tensors}. $Ch_{BE}$ is a tunable parameter. Memory-bound tensors are identified through offline profiling (\S\ref{sec:vram-bandwidth-allocation}). 

ii) \textbf{Intra-SM conflicts.} SGDRC schedules LS and BE tasks based on spatial-temporal multiplexing to prevent intra-SM conflicts. At any given time, only one LS kernel and one BE kernel can be colocated on the GPU. LS (or BE) kernels from different tasks are launched to the LS (or BE) kernel queue in a round-robin manner. SGDRC enables LS tasks to preempt compute resources occupied by BE tasks and dynamically allocates compute units using \textit{tidal SM masking} (Fig.~\ref{fig:spatial-temporal-multiplexing-illustrate}a$\sim$c). As kernels running on different GPU models have diverse resource sensitivities, SGDRC allocates $SM_{LS}$ TPCs to each LS kernel, which is the minimum number of TPCs required to achieve the lowest latency for LS kernels. This parameter is also determined by offline profiling.

%\subsection{\textsc{Missile}'s Workflow}

\vspace{-0.1in}

%\begin{figure}
%	\centering
%	\includegraphics[width=0.8\linewidth]{sections/inter-sm-contention/figures/resource-curve.pdf}
%	\caption{SGDRC's SM (a) and VRAM bandwidth (b) allocation curves, which guide the dynamic SM and VRAM channel allocation based on GPU kernels' profiling results.}
%	\label{fig:allocation-curve}
%\end{figure}

 %During compilation, it also uses a simple greedy algorithm to generate an efficient execution plan (e.g., when to load and reclaim tensors, and execute kernels).

%2) \textbf{Online phase.} The DNN's parameters, pre-compiled CUDA kernels, and the execution plan collectively form an \textit{inference task}. \todo{TODO.}%The global memory space and VRAM channels are divided into two segments, with each division being shared by all LS and BE tasks, respectively.} %\revised{Given that BE tasks consume more VRAM-bandwidth and VRAM space}, we \revised{empirically} assign 1/3 \revised{ and 2/3} VRAM channels \revised{(and global memory space)} to LS and BE tasks\revised{, respectively}.%All \texttt{cuMemcpy} operations from each task are forwarded to the memory copy queues, and the completely fair scheduler is responsible for serially selecting and executing memory copy requests from these queues. \textsc{Missile} is capable of executing multiple LS models concurrently. When multiple BE models are colocated on the same GPU, to maximize the cache and VRAM bandwidth utilization, at each time, at most one BE kernel can be colocated with multiple LS kernels. BE models launch their kernels in a round-robin manner. 

\section{Reverse Engineering VRAM Channels}\label{sec:vram-channel-reverse-engineering}

\subsection{Labeling VRAM Channel IDs}\label{subsec:label-vram-channels}

Before reverse engineering VRAM channels, we need to establish the mapping between a program's virtual VRAM addresses and GPU physical addresses, because a virtual VRAM space is randomly mapped to a part of the physical VRAM space and thus the mapping between virtual VRAM addresses and VRAM channel IDs changes each time the program restarts. We follow the practice of \cite{Zhang2023} to fetch physical addresses by parsing the page table entries stored in the VRAM.  

After that, we label VRAM channel IDs in the physical VRAM space. As GPUs adopt the UMA (Uniform Memory Access) architecture, we cannot identify VRAM channels by measuring the cache access latency, as commonly done on CPUs. We observe that any pair of physical addresses with a DRAM bank conflict or an L2 cacheline conflict must belong to the same VRAM channel because a DRAM bank or L2 cacheline is associated with only one VRAM channel (as demonstrated in \S\ref{subsec:nvidia-gpu-arch}). Thus, we can identify all addresses that reside in the same channel by populating all available L2 cachelines in the channel, which requires two steps:

1) \textbf{Generating a set of conflicted addresses belonging to \texttt{Addr}'s VRAM channel.} We find a series of addresses \texttt{DramConflictAddrs} that have DRAM bank conflicts with \texttt{Addr}. This is achieved by concurrently reading from \texttt{Addr} and \texttt{Addr'} and measuring the access latency \revised{(Algo.~\ref{algo:find-dram-bank-conflict} in \S\ref{appendix:reverse-engineering-algorithm})}. Then, we retrieve a series of addresses, \texttt{CacheConflictAddrs}, that have cache conflicts with \texttt{DramConflictAddrs} by binary searching the minimum interval $($\texttt{Addr},  \texttt{Addr'}$]$ that can evict \texttt{Addr} from the L2 cache (Algo.~\ref{algo:find-cacheline-conflict} in \S\ref{appendix:reverse-engineering-algorithm}).

2) \textbf{Identifying the VRAM channel ID of a given address.} After generating a set of addresses belonging to each VRAM channel, we can now identify the VRAM channel ID to which any given address \texttt{Addr'} is mapped. This process involves three steps: a) Reading \texttt{Addr'} to populate it into a cacheline; b) Reading \texttt{CacheConflictAddrs} belonging to the $i$-th VRAM channel to refresh all cachelines in the $i$-th VRAM channel; and c) Reading \texttt{Addr'} again and timing its latency. If the latency exceeds the threshold (determined by micro-benchmarking~\cite{Mei2017}), it indicates an L2 cache miss, and thus \texttt{Addr} maps to the $i$-th VRAM channel (Algo.~\ref{algo:mark-memory-channel} \revised{in \S\ref{appendix:reverse-engineering-algorithm}}).

\subsection{Findings of VRAM Channel Mapping Structure}\label{subsec:findings-vram-channel}

\begin{figure}[tb]
	\centering
	\includegraphics[width=\columnwidth]{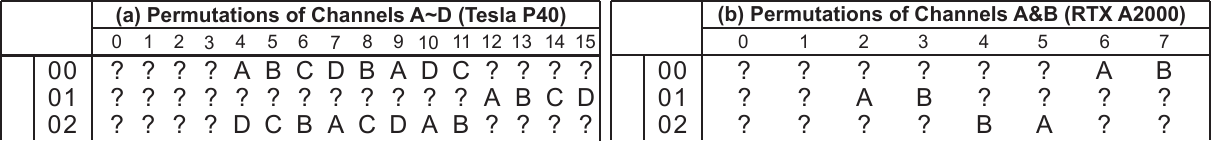}
	\caption{A part of VRAM channel permutations of Tesla P40 (channels \texttt{A} $\sim$ \texttt{D}) and RTX A2000 (channels \texttt{A} $\sim$ \texttt{B}). \textbf{Note:} 1) \texttt{?} denotes VRAM channels not in \texttt{A} $\sim$ \texttt{D} (\texttt{A} $\sim$ \texttt{B}); 2) Full results can be found in Fig.\ref{fig:full-reverse-engineering-results} of \S\ref{appendix:reverse-engineering-findings}.}
	\label{fig:simplified-reverse-engineering-results}
\end{figure}

In a contiguous 10 MiB VRAM space, we mark VRAM channels for both the Tesla P40 and the RTX A2000. The marking results (Fig.\ref{fig:simplified-reverse-engineering-results} and Fig.\ref{fig:full-reverse-engineering-results}) show that each contiguous 1 KiB of physical VRAM space belongs to the same \revised{VRAM} channel (i.e., a \textit{channel partition}). The mapped VRAM channel IDs of contiguous \textit{channel partitions} form an \textit{m-permutation} ($m$ is the number of patterns in the permutation). For the Tesla P40, channels A$\sim$D, E$\sim$H, and I$\sim$L form 3 independently distributed 24-permutations. For the RTX A2000, channels A$\sim$B, C$\sim$D, and E$\sim$F form 3 independently distributed 12-permutations.

\begin{figure}
	\centering
	\includegraphics[width=0.9\linewidth]{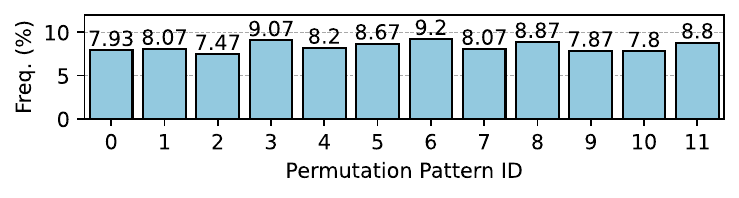}
	\caption{The frequency histogram of RTX A2000's 12 permutation patterns of VRAM channels A \& B. All patterns are uniformly distributed across the 12 GiB VRAM space.}
	\label{fig:3090permutationfrequencyhist}
\end{figure}

\begin{figure}
	\centering
	\includegraphics[width=1\columnwidth]{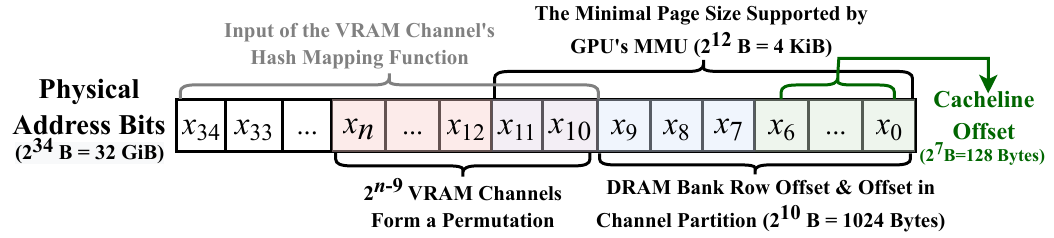}
	\caption{Structure of NVIDIA GPU's physical address bits.}
	\label{fig:nvidia-gpu-phys-addr-bit}
\end{figure}

All permutation patterns are uniformly distributed (Fig.~\ref{fig:3090permutationfrequencyhist}), and the occurrence frequency of each VRAM channel ID among all permutation patterns is equal. This indicates that VRAM channels are evenly distributed across the VRAM space. We summarize the physical address structure in Fig.~\ref{fig:nvidia-gpu-phys-addr-bit}, where bits 10 $\sim$ 34 serve as the input to the black-box hash mapping function that generates the VRAM channel ID assigned to a physical address. In each permutation of Tesla P40 (or RTX A2000), at most 4 KiB (or 2 KiB) space shares the same set of VRAM channels.

\subsection{Cracking the VRAM Channel's Hash Mapping}\label{subsec:fit-hash-mapping}

Most GPUs' VRAM channel hash functions are not purely XOR-based and thus cannot be cracked directly (\S\ref{subsec:hardware_patitioning_limitations}), but marking each address's channel ID in the entire VRAM space is also extremely time-consuming. For example, if the VRAM size is 24 GiB, it would require marking 24 GiB/1024 B = 25 million VRAM channels, which would take more than 1 year to complete. Therefore, we need to crack the hash mapping of VRAM channels using a non-brute-force solution. Although the mapping functions of VRAM channels in many NVIDIA GPUs are non-linear, and the structure of these functions is unknown, fortunately, we can use DNNs to offline approximate them: DNNs have already been proven to be theoretically capable of statistically meaningful approximation of any boolean function~\cite{Wei2022}. For each GPU model, we spend one month collecting 15K samples of VRAM channel mapping and training a DNN to fit this mapping function. We then spend 1 hour using this DNN to make inferences in batch and offline generate a \textit{lookup table}, which stores the VRAM channel ID of each 1 KiB channel partition across the VRAM space. The results on the test set indicate that our DNN can accurately label over 99.9\% of VRAM channel IDs when provided with an unseen physical address.

\begin{figure}
	\centering
	\includegraphics[width=1\columnwidth]{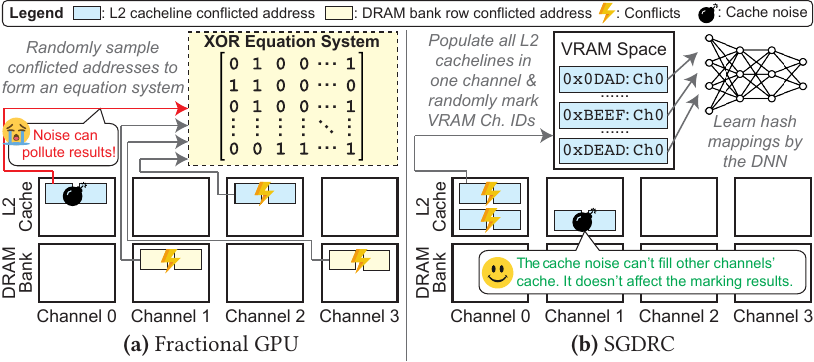}
	\caption{Illustrative comparison between FGPU's~\cite{Jain2019} and SGDRC's reverse-engineering approaches.}
	\label{fig:reverse-engineering-compare}
	\vspace{-0.1in}
\end{figure}

We compare FGPU's and SGDRC's reverse engineering approaches in Fig.~\ref{fig:reverse-engineering-compare}. In our practice, only around 5\% of the conflicted addresses obtained in step 1) of \S\ref{subsec:label-vram-channels} are mapped to other channels due to the cache noise. Once enough conflicted addresses are populated into the L2 cache, only the target VRAM channel will be filled. Therefore, this approach can tolerate the cache noise.

\section{Enabling VRAM Bandwidth Partition}\label{sec:vram-bandwidth-allocation}

Once the VRAM channel mapping is available, SGDRC utilizes cache coloring to isolate VRAM channel conflicts between LS and BE kernels. Since the physical address space belonging to the same VRAM channel is not contiguous, we need to remap the address space accessed by each task to the given set of VRAM channels. An intuitive idea is to intercept the VRAM allocation in \texttt{nvidia-uvm} and allocate pages with the same color to the given task, as adopted by FGPU~\cite{Jain2019}. However, it only supports 4 KiB coloring granularity, which is the minimal page size supported by the GPU's MMU.

However, our reverse engineering results (\S\ref{subsec:findings-vram-channel}) have revealed limitations to this approach. In the VRAM channel layout of the RTX A2000, the VRAM space is composed of a series of paired VRAM channels, which means that on new GPU architectures, the coloring granularity can only be 1 KiB or 2 KiB, and larger values are inapplicable (we discuss how to decide this in \S\ref{appendix:decide-coloring-granularity}). Thus, enabling cache coloring with an $n$ KiB size ($n$=1 or 2) must involve overheads brought by extra address remapping, and the limited GPU register size makes this problem more challenging.

\begin{figure}[tb]
	\centering
	\vspace{-.15in}
	\subfloat[Driver-level page mapping. The coloring granularity is 2 KiB. \textbf{Note:} \texttt{VPN}: Virtual Page Frame Number; \texttt{PFN}: Physical Page Frame Number.]
	{
		\includegraphics[width=1\columnwidth]{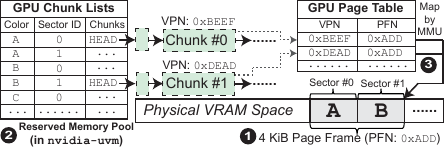}
		\label{fig:demo_shadow_page_table}
	}\\
	\subfloat[The original BE kernel in \textbf{(a)}, which performs vector addition.]
	{
		\includegraphics[width=1\columnwidth]{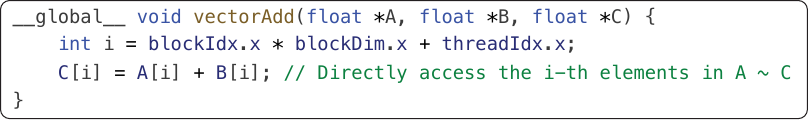}
		\label{fig:original-vec-add}
	}\\
	\subfloat[The transformed DNN kernel of BE task in \textbf{(a)}. Virtual addresses of each colored tensor should be adjusted by adding \textit{the sector index} $\times$ \textit{sector size}.]
	{
		\includegraphics[width=1\columnwidth]{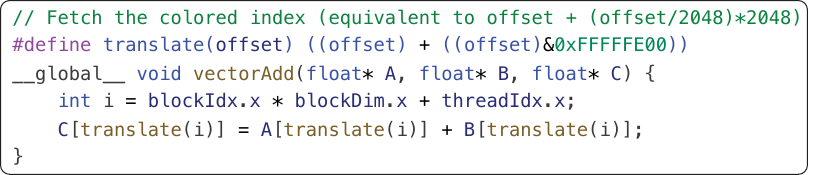}
		\label{fig:transformed-vec-add}
	}
	% \vspace{-.15in}
	\caption{Illustration of the shadow page table (SPT).}
	\vspace{-.1in}
	\label{fig:dnn_kernel_transformation_example}
\end{figure}

SGDRC binds tasks to their corresponding VRAM channels by introducing the \textit{shadow page table} (SPT, Fig.~\ref{fig:dnn_kernel_transformation_example}), which includes: 1) Dividing each 4 KiB page into $\frac{4}{n}$ sectors with IDs from 0 to $n$-1 (Fig.~\ref{fig:demo_shadow_page_table} \ding{182}) and marking each sector's color using the lookup table (\S\ref{subsec:fit-hash-mapping}); 2) Reserving a memory pool in the \texttt{nvidia-uvm} kernel module and maintaining linked lists for $n$ KiB chunks with different colors (Fig.~\ref{fig:demo_shadow_page_table} \ding{183}); 3) Writing the physical page frame number of each chunk to the GPU's page table (Fig.~\ref{fig:demo_shadow_page_table} \ding{184}); and 4) Transforming array indexes in DNN kernels to remap tensors to $n$ KiB sectors with the same color and chunk ID (each re-indexing operation requires 2 integer operations, 8 GPU cycles, Fig.~\ref{fig:original-vec-add}-\ref{fig:transformed-vec-add}). The addresses of arguments passed into the kernels are also adjusted by adding \textit{the sector index} $\times$ \textit{sector size} for alignments. SGDRC allocates $Ch_{BE}$ and $(1-Ch_{BE})$ of VRAM channels to BE and LS tasks, respectively. The coloring granularity and $Ch_{BE}$ are set to be 2 KiB and 1/3. Coloring granularity and $Ch_{BE}$ are tunable. However, there are only a few valid values for Tesla P40 and RTX A2000. So SGDRC chooses not to tune them. We leave this problem as a future work for GPUs with more VRAM channels.

\section{Dynamic Resource Allocation}\label{sec:elastic-sm-allocation}

\subsection{Elastic SM Scaling}
\begin{figure}
	\centering
	\includegraphics[width=1.0\linewidth]{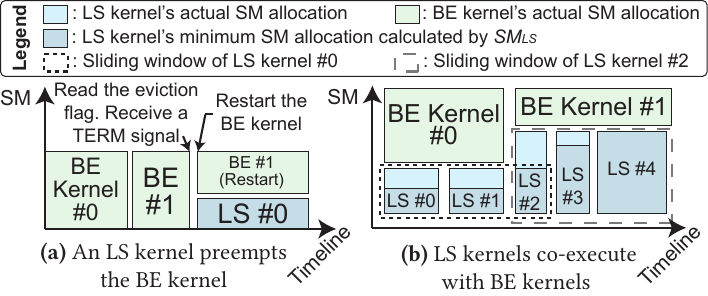}
	\vspace{-0.3in}
	\caption{Illustration of the \textit{tidal SM masking}.}
	\label{fig:tidal-sm-masking-demo}
\end{figure}

SGDRC uses \texttt{libsmctrl}~\cite{Bakita2023}, a library that manipulates Task Meta Data (TMD~\cite{tmd-patent}, an NVIDIA-specific, little-known interface), to control the set of TPCs to which each launched kernel can be assigned. LS kernels can preempt SMs occupied by BE kernels following the designs of FLEP~\cite{Wu2017} and Reef~\cite{Han2022}. Specifically, SGDRC checks an eviction flag stored in global memory using \texttt{ld.cv}. LS tasks can write the eviction flag to preempt a BE kernel. After that, the BE kernel restarts and co-executes with LS kernels (Fig.~\ref{fig:tidal-sm-masking-demo}a).

When LS and BE kernels are co-executed, SGDRC elastically allocates SMs to LS and BE kernels to maximize the SM unit utilization (Fig.~\ref{fig:tidal-sm-masking-demo}b). SGDRC uses a binary search during offline profiling to determine the minimum number of SMs required by each LS kernel to achieve optimal latency. Given that the runtime of BE kernels may be longer than that of LS kernels, and that LS kernels waiting in the kernel launch queue may consume more SMs than the currently allocated ones, SGDRC determines the actual SM allocation based on a sliding window. The number of SMs reserved for the next LS kernel is the maximum number of SMs required by LS kernels in the sliding window. SGDRC also transforms LS and BE kernels with a large number of thread blocks into the persistent-thread style to reduce conflicts caused by the GPU hardware scheduler. More details about the persistent-thread-style GPU programming can be found in the implementation described in \cite{Wu2015}.

\subsection{Dynamic VRAM Channel Allocation}\label{subsec:dynamic-bandwidth-allocation}

\begin{figure}
	\centering
	\vspace{-0.1in}
	\includegraphics[width=1.0\linewidth]{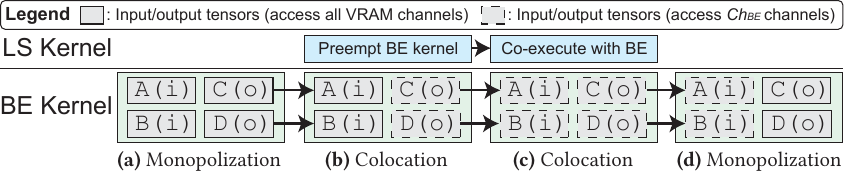}
	\vspace{-0.3in}
	\caption{Illustration of \textit{bimodal tensors}. \textbf{Note:} \texttt{(i)}: Input tensor; \texttt{(o)}: Output tensor.}
	\label{fig:transition-kernel-demo}
	\vspace{-0.2in}
\end{figure}

SGDRC uses \textit{bimodal tensors} to enable dynamic VRAM bandwidth allocation (Fig.~\ref{fig:transition-kernel-demo}). To reduce the overhead, SGDRC identifies (through offline profiling) and isolates \textit{memory-bound} tensors (tensors accessed by \textit{memory-bound} kernels). A kernel is considered \textit{memory-bound} if its runtime degrades when L2 cachelines are intensively populated by a colocated kernel. Each memory-bound LS tensor is mapped to $(1-Ch_{BE})$ of the VRAM channels. When no BE task is colocated, LS tasks can fully utilize the VRAM bandwidth by moving all memory-bound tensors to the reserved memory pool mapped to all VRAM channels. To achieve fast VRAM bandwidth scaling, for each memory-bound BE tensor, SGDRC maintains its 2 copies : 1) one mapped to all VRAM channels; and 2) one mapped to $Ch_{BE}$ of the VRAM channels. For each memory-bound BE kernel, SGDRC passes its parameters based on two cases (Fig.~\ref{fig:transition-kernel-demo}):

a) \textit{Monopolization state} (when the LS kernel queue is empty): All input weight tensors and output tensors are mapped to all VRAM channels.

b) \textit{Colocation state} (when LS kernels and BE kernels are colocated): Memory-bound input weight tensors and memory-bound output tensors are mapped to $Ch_{BE}$ of the VRAM channels.

For both cases and non-memory-bound kernels, SGDRC decides whether input intermediate tensors should be mapped to $Ch_{BE}$ of the VRAM channels based on the state of the last kernel that accesses these tensors. To further minimize extra memory usage introduced by tensor copies, SGDRC fully reuses tensors storing intermediate results.

\section{Implementation}\label{sec:implementation}

\revised{SGDRC is implemented in C++ with \textasciitilde12K LOC (\textasciitilde1K LOC for reverse engineering, \textasciitilde2K LOC for cache coloring in \texttt{nvidia-uvm}, \textasciitilde2K LOC for the kernel transformer, and \textasciitilde7K LOC for the inference server and client).} It utilizes TVM~\cite{Chen2018} and Ansor~\cite{Zheng2020} to generate and optimize CUDA kernels.

\section{Evaluation}\label{sec:final-evaluation}

Having discussed how SGDRC reverse engineers VRAM channel mapping (\S\ref{sec:vram-channel-reverse-engineering}), partitions VRAM channels (\S\ref{sec:vram-bandwidth-allocation}), dynamically allocates resources (\S\ref{sec:elastic-sm-allocation}), and is implemented (\S\ref{sec:implementation}), we now evaluate it to answer the following key questions:

1) What are the gains and overheads of VRAM bandwidth partitioning and dynamic resource allocation? (\S\ref{subsec:micro-benchmarking})

2) Can SGDRC effectively mitigate resource contention among tasks and enable dynamic resource allocation?  (\S\ref{subsec:final_evaluation_results})

\subsection{SGDRC Performance Deep Dive}\label{subsec:micro-benchmarking}

We begin by quantifying the performance gains and overheads of \textit{shadow page tables} and \textit{bimodal tensors}.

\subsubsection{VRAM channel isolation performance gains.} We conduct tests on both Tesla P40 and RTX A2000, using NVIDIA Nsight Compute to profile kernels. We randomly select some kernels with high DRAM throughput from BE models to act as the source of VRAM channel conflicts. All kernels from LS models listed in Tab.~\ref{tab:list_of_testing_models} are incorporated to evaluate the extent of interference caused by memory-bound BE kernels and the overhead introduced by the shadow page tables. 

\begin{figure}
	\centering
	\includegraphics[width=\columnwidth]{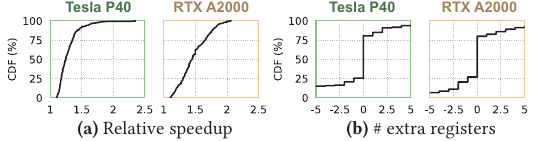}
	\vspace{-0.3in}
	\caption{Evaluation of VRAM channel isolation. (a) CDF of LS kernels' runtime speedup after applying VRAM channel isolation. Larger values are better; (b) CDF of extra registers used by VRAM channel isolation (the \# of registers used by the transformed kernel minus the \# of registers used by the original kernel). Smaller values are better.} %(c) VRAM footprint of 3 BE models, smaller values are better.}%A memory-intensive BE kernel is coexecuted with LS kernels. Larger values are better.}
	\label{fig:shadow_page_table_isolation_cdf_by_kernel}
	\vspace{-0.2in}
\end{figure}

The experimental group allocates $1-Ch_{BE}$ and $Ch_{BE}$ of the VRAM channels to the memory-intensive tensors of LS and BE kernels, respectively (in our setting, $Ch_{BE}$ is tuned to 1/3). Then we coexecute each LS kernel with the selected memory-intensive BE kernels in a closed loop. The control group does not enable VRAM channel isolation. Both the experimental group and the control group utilize libsmctrl~\cite{Bakita2023} to evenly partition SMs for LS and BE kernels. We compare the p99 latency of the LS kernel runtime between the experimental group and the control group (Fig.~\ref{fig:shadow_page_table_isolation_cdf_by_kernel}a). The results reveal that for all LS kernels, on Tesla P40 and RTX A2000, SGDRC's VRAM channel isolation reduces p99 latencies compared with the non-isolated control group by 28.7\% and 47.5\% on average and by up to 135\% and 106.3\%, respectively. Note that these results account for incorrect VRAM channel predictions, as mispredicted VRAM channel IDs are randomly distributed across the VRAM space.

\subsubsection{Overheads of VRAM channel isolation.}

Although SGDRC's VRAM channel isolation exhibits outstanding performance, the overhead introduced by SPTs remains a focal point of concern, as it incurs extra register usage and calculations for array re-indexing. Theoretically, each thread requires only one additional register to store the intermediate value for array re-indexing. For real-world kernels, we use \texttt{nvcc -O3} to compile kernels and compare the register usage between their transformed and original implementations. On Tesla P40 and RTX A2000, 80.4\% and 80.0\% kernels do not use extra registers; 93.8\% and 91.2\% kernels use fewer than 5 extra registers. We observe that a few transformed kernels use more than 10 extra registers. Upon inspection, these outliers are small kernels with runtime shorter than 0.01ms. Their register usage is influenced by \texttt{nvcc} compiler optimizations and has little impact on a DNN's end-to-end performance.

Next, we allocate all VRAM channels to each DNN with SPT. In the absence of colocated BE kernels, we compare the p99 runtime of transformed kernels with that of the original kernels. Across all DNN kernels on Tesla P40 and RTX A2000, the overhead of SPT is 2.9\% on average. After applying SPTs to memory-bound kernels, the end-to-end DNN inference time (including CPU-side operations) increases by $\sim$0.5\% on average.

\subsubsection{VRAM footprints of bimodal tensors}

\begin{figure}
	\centering
	\includegraphics[width=1.0\columnwidth]{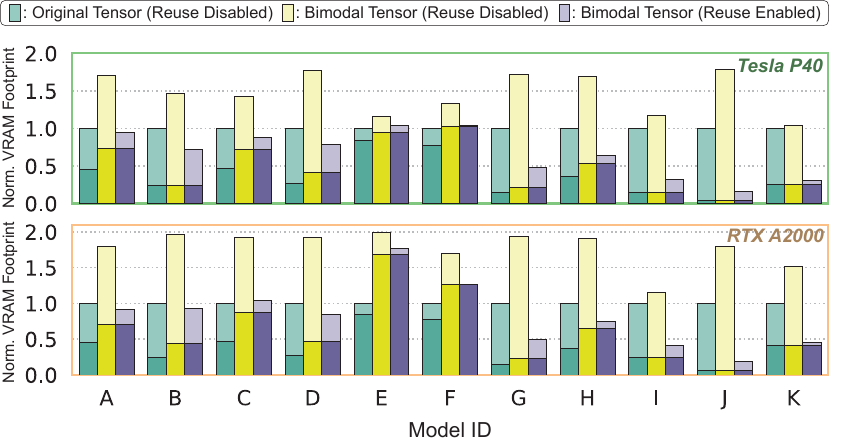}
	\vspace{-0.3in}
	\caption{VRAM footprints introduced by bimodal tensors. Dark colors denote weight tensors, and light colors denote intermediate tensors.} %(c) VRAM footprint of 3 BE models, smaller values are better.}%A memory-intensive BE kernel is coexecuted with LS kernels. Larger values are better.}
\label{fig:vram-footprint-bimodal-tensors}
\vspace{-0.1in}
\end{figure}

We measure the VRAM footprints of \textit{bimodal tensors} in Fig.~\ref{fig:vram-footprint-bimodal-tensors}. Without reusing the intermediate tensors, the VRAM footprints of all DNNs nearly double. Reusing intermediate tensors (\S\ref{subsec:dynamic-bandwidth-allocation}) can significantly reduce the VRAM footprints, especially in BE models I $\sim$ K, because they have larger batch sizes than LS models and require more space to store intermediate results.

%\subsubsection{Reconfiguration overheads.}

\subsection{\revised{End-to-End Experimental Setup}}\label{subsec:final_experimental_setup}

\PHM{Testbeds.} We deploy and evaluate SGDRC on two GPUs: (a) Tesla P40 (representing deprecated GPU architectures) and (b) RTX A2000 (representing new GPU architectures). To mitigate the impact of network latency, we follow Reef~\cite{Han2022} and deploy SGDRC's server and clients on the same machine.

\PHM{Workloads.} We reference \cite{Ma2020,Zhu2022,Han2022,Ng2023,Shi2023,Wu2023,Foteini2024} and select \revised{12} representative LS and BE models as testing workloads (Tab.~\ref{tab:list_of_testing_models}). To minimize the interference between LS and BE tasks, we set the batch sizes of BE tasks to be the minimum values that achieve maximum throughputs. LS services' clients send requests by replaying Baidu's Apollo trace~\cite{disb_benchmark}, a real-time DNN inference trace collected from Baidu's Apollo autonomous driving system. \revised{Both were} adopted by Reef~\cite{Han2022} and Orion~\cite{Foteini2024} to evaluate their systems. BE tasks run in a closed-loop manner.

\begin{table}[]
	\centering
	\caption{Testing DNN models.}
	\scriptsize
	\begin{tabular}{ccc}
		\toprule
		\textbf{\revised{Class}} & \textbf{Model Name} \\
		\midrule
		\textbf{LS}       & \makecell[c]{MobileNetV3 (A), SqueezeNet (B), ShuffleNet (C), EfficientNet (D),\\ResNet34 (E), \revised{MobileBert (F)}, \revised{MobileViT (G),} \revised{EfficientFormer (H)}}           \\
		\hline
		\textbf{BE}       & \makecell[c]{ResNet152 (\revised{I}), DenseNet161 (\revised{J}), \revised{Bert (\revised{K})}}       \\
		\bottomrule
	\end{tabular}
	\label{tab:list_of_testing_models}
\end{table}

\PHM{Testing scenarios.} On two GPUs, we deploy LS models A $\sim$ H and BE models I $\sim$ K in GPUs' VRAM before serving requests. Given that dynamic batching is detrimental to request latency as it requires early requests to wait for batching with additional requests~\cite{Ng2023}, we don't incorporate this feature into SGDRC and the baselines. To serve multiple requests concurrently, each LS model has 4 instances. All LS services are simultaneously colocated with a BE task at all times. We evaluate all systems in two scenarios to measure their performance under varying workloads: 1) \textit{Light workload}: The Apollo trace is scaled to reduce the average request rate to half of its original value; 2) \textit{Heavy workload}: Use the original Apollo trace.

\PHM{Baselines.} We select the following solutions as baselines:

1) \textit{Multi-streaming}: To reduce interference, we deploy two kernel launch streams (LS \& BE) in \textit{Multi-streaming} and assign a higher stream priority to the LS queue. Requests from LS and BE tasks are forwarded to these streams in a round-robin manner. 

2) \textit{TGS}~\cite{Wu2023}: Since TGS inherently supports the colocation of only one LS and one BE container, we forward requests from each LS service and BE task to these two containers in a round-robin manner.

3) \textit{MPS}~\cite{nvidia-mps}: As the maximum number of instances supported by MPS is constrained, and too many MPS instances concurrently executing on one GPU can lead to severe contention, we evenly divide the GPU into two MPS instances, use \texttt{CUDA\_MPS\_ACTIVE\_THREAD\_PERCENTAGE} to limit the compute resource usage of each instance, and serve LS and BE tasks on them separately in a round-robin manner. 

4) \textit{Orion}~\cite{Foteini2024}: Considering that Orion's code only supports DNN inference on PyTorch's backend with outdated cuDNN libraries, we implement Orion's scheduling policy within SGDRC to ensure a fair comparison.

5) \textit{SGDRC (Static)}: A vanilla version of SGDRC that statically and evenly partitions resources between LS and BE tasks.

\PHM{Evaluation metrics.} For LS services, we gather their p99 latency \revised{(including queueing delays)} and the SLO attainment rates. We follow the settings of \cite{Choi2021,Chow2023} and set the SLO to be $n \times$ p99 isolated execution runtime of each DNN model (where $n$ is the number of DNN services concurrently running on the GPU). For BE tasks, we collect their throughput (number of samples processed per second). We also measure the overall throughput (LS services' goodput + BE tasks' throughput). Since BE tasks are co-located with LS services in a round-robin manner, we record both the overall throughput and the throughput per BE task to better understand each system's performance.

\begin{figure*}[t]
	\centering
	\includegraphics[width=1\linewidth]{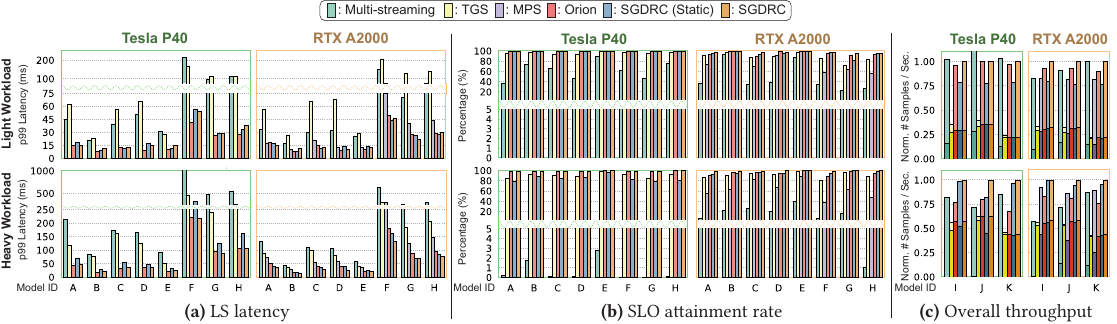}
	\caption{End-to-end evaluation. \revised{\textbf{Note:}} MPS \revised{is no longer supported on} P40; Dark colors in \textbf{(c)} denote the LS service goodput.}
	\label{fig:finalevalfirstpart}
\end{figure*}

\subsection{Evaluation Results}\label{subsec:final_evaluation_results}

%\begin{figure}[t]
%	\begin{minipage}[t]{0.48\textwidth}
%		\centering
%		\includegraphics[height=2.5cm]{sections/evaluation/figures/ablation_study}
%		\caption{Ablation study.}
%		\label{fig:ablation_study}
%	\end{minipage}
%\end{figure}

We present the results in Fig.\ref{fig:finalevalfirstpart}. In both \textit{light} and \textit{heavy workload} scenarios, SGDRC achieves the highest SLO attainment rate (99.0\% on average). It demonstrates low p99 latency (comparable to or lower than Orion) for all LS models. Its throughput is lower than Multi-streaming in some cases. But Multi-streaming sacrifices LS services' tail latency and thus have low LS service goodput. TGS exhibits both high p99 latency for LS services and the lowest throughput. This can be attributed to: 1) the substantial overhead resulting from frequent CUDA context switches between GPU containers, and 2) the feedback-based dynamic sending rate control algorithm, which fluctuates containers' resource allocation. The undesirable LS p99 latency and SLO attainment rates of MPS can be attributed to the fact that MPS isolates SM resources at thread level without addressing intra-SM and VRAM channel conflicts. Across Tesla P40 and RTX A2000, compared to the state-of-the-art solution (Orion), SGDRC improves the overall throughput by up to 1.47$\times$ and 1.32$\times$, BE task throughput by up to 2.36$\times$ and 1.79$\times$, respectively. This significant improvement can be attributed to the challenges faced by Orion in identifying suitable co-executed BE kernels (as elaborated in \S\ref{subsec:limitation_existing_gpu_sharing}). Compared with SGDRC (Static), SGDRC achieves higher BE job throughput, which is more evident in the \textit{light workload} scenario. This can be accredited to the dynamic resource management policy, which maximizes the GPU resource utilization.

\section{Discussion and Future Work}\label{sec:discussion}

\PHM{Integration with NVIDIA MIG.} The objective of SGDRC is to offer hardware resource isolation for DNN inference services on mid-to-low-end GPUs that lack MIG support. In the future, SGDRC may integrate with MIG on NVIDIA A100 and H100, enabling dynamic resource partitioning within each MIG instance. However, it is crucial to acknowledge that the L2 caches in A100~\cite{nvidia-a100-whitepaper} and H100~\cite{nvidia-h100-whitepaper} consist of multiple separate caches, making their L2 caches a hybrid of UMA and NUMA. Consequently, SGDRC's reverse engineering approach will require slight adaptation for NVIDIA A100 and H100.

\PHM{Extending to transparent task colocation.} Like Clockwork~\cite{Gujarati2020}, Reef~\cite{Han2022}, Paella~\cite{Ng2023}, and StreamBox~\cite{StreamBox}, SGDRC utilizes TVM~\cite{Chen2018} to generate CUDA kernels for DNN inference tasks. However, CUDA kernels of other GPU tasks may originate from closed-source vendor libraries (e.g., cuDNN, Cutlass, and cuBLAS) and CUDA binaries, which are not currently compatible with SGDRC. Previous research has demonstrated the possibility of modifying CUDA kernel code from closed-source libraries and binaries through GPU program disassembly~\cite{Zhou2015,Mai2023,Pavlidakis2024}. Therefore, SGDRC has the potential to support colocating heterogeneous tasks in the future.

\PHM{Fault isolation.} Like Reef~\cite{Han2022}\revised{, Paella~\cite{Ng2023}, and Orion}~\cite{Foteini2024}, SGDRC aggregates multiple workloads into one CUDA context to reduce task-switching overhead. However, SGDRC and these solutions cannot isolate colocated DNNs' GPU runtime errors. We believe this is acceptable, as SGDRC relies on TVM~\cite{Chen2018} to generate and check CUDA kernels, and runtime errors due to incorrect kernel implementations are unlikely to occur. \revised{In the future, SGDRC could isolate faults through static analysis-based~\cite{Mai2023} or runtime-based~\cite{Pavlidakis2024} methods.}

\PHM{Supporting serverless ML services.} Existing work~\cite{Hui2024} leverages NVIDIA MIG~\cite{nvidia-mig} to eliminate resource contention on serverless ML platforms. As SGDRC provides dynamic resource allocation for low-end GPUs, it could also be adapted to serverless ML infrastructures in the future.

\section{Conclusion}

This paper presents SGDRC, a general, software-defined GPU dynamic resource control solution for concurrent DNN inference on NVIDIA GPUs. SGDRC isolates inter- and intra-SM conflicts through full-spectrum reverse engineering of GPU VRAM channels, learning VRAM channel hash mapping by DNNs, fine-grained software-level cache coloring, and dynamic VRAM channel and compute unit allocation. SGDRC mitigates the degradation of BE throughput observed in other GPU sharing techniques with conservative policies. Compared with state-of-the-art GPU sharing solutions, SGDRC achieves the highest SLO attainment rates (99.0\% on average) and improves overall throughput by up to 1.47$\times$ and BE job throughput by up to 2.36$\times$.

\begin{acks}
	We thank our anonymous reviewers for their valuable feedback. We thank Dr. Liping Zhang (Alibaba Group), Yuanchi Zhang (Tsinghua University), Guoqing Luo (University of Alberta), Junliang Hu (CUHK), and Guangming Sheng (HKU) for their helpful insights at various stages of the project. Yongkang Zhang is supported by the Hong Kong Ph.D. Fellowship Scheme (PF20-46117). Dr. Huaicheng Li is supported by the National Science Foundation under Grant CNS-2312785 and the CAREER Award CNS-2339901.
\end{acks}

\appendix

\section{Appendix}

\subsection{VRAM Channel Reverse Engineering Algorithms}
\label{appendix:reverse-engineering-algorithm}
Here, we describe the algorithms used for reverse engineering the VRAM channel hash mapping, including: 1) Identifying a series of physical addresses that have DRAM bank conflicts with a given physical address (Algo.~\ref{algo:find-dram-bank-conflict}); 2) Identifying a series of physical addresses that experience L2 cacheline conflicts with a given physical address (Algo.~\ref{algo:find-cacheline-conflict}); and 3) Assigning a VRAM channel ID to a given physical address (Algo.~\ref{algo:mark-memory-channel}). They use the GPU pointer-chase (P-chase) algorithm to populate the L2 cache and detect L2 cacheline conflicts. More details about the P-chase algorithm can be found in \cite{Mei2017}.

\begin{algorithm}[tb]
	\caption{Find DRAM bank conflict addresses}
	\footnotesize
	\begin{algorithmic}
			\Function{IsDramBankConflicted}{Addr0, Addr1}
			\State $v \gets [0,1,2,3,......]$ \Comment{Initialize the pointer chase array $v$}
			\State $RefreshL2(v)$ \Comment{Use P-chase to refresh the L2 cache}
			\State $StartTime \gets Clock()$
			\State $Addr0 \gets v[Addr0]$ %\Comment{Read $Addr0$ and $Addr1$ from the DRAM together.}
			\State $Addr1 \gets v[Addr1]$
			\State $EndTime \gets Clock()$
			\If{$EndTime - StartTime > THRESHOLD$}
			\State \Return True \Comment{Indicate the DRAM bank conflict}
			\Else
			\State \Return False
			\EndIf
			\EndFunction
		\end{algorithmic}
	\label{algo:find-dram-bank-conflict}
\end{algorithm}

%\subsection{Detecting L2 Cacheline Conflict Addresses}\label{subsec:find_cacheline_conflict}

\begin{algorithm}[tb]
	\caption{Find L2 cacheline conflict addresses}
	\footnotesize
	\begin{algorithmic}
	%\COMMENT{Test if reading data in [Addr0, Addr1] can cause the cacheline eviction}
	
			\Function{IsCachelineEvicted}{array, Addr0, Addr1}
	
			\State{Pointer-chase[$array[Addr0:Addr1]$} \Comment{Use P-chase to read the interval [$Addr0$, $Addr1$] and populate the L2 cache.}
	
			\State $t = TimeElapse(addr \gets array[Addr0])$ \Comment{Measure the latency of re-accessing the element $array[Addr0]$}
	
			\If{$t > Thres_{L2\ Miss}$} \Comment{Indicate an L2 cache miss}
				\State \Return True
			\Else  \Comment{Indicate an L2 cache hit}
				\State \Return False
			\EndIf
			\EndFunction
		\end{algorithmic}
	\begin{algorithmic}
			\Function{FindCacheConflictAddrs}{Addr}
			
			\State $array \gets [0,1,2,3,......]$ \Comment{Initialize the P-chase array}
			\State $ConflictAddrList \gets []$
			\For{$i \gets 0$ to $MAX\_ITER$}
				\State $lower\_bound \gets 1$
				\State $upper\_bound \gets MAX\_UPPER\_BOUND$ %\Comment{The next L2-cacheline-conflicting address should be in the range [$Addr+lower\_bound$, $Addr+upper\_bound$]}
				\State $ConflictAddr \gets Addr$
				\While{$lower\_bound < upper\_bound$}
					\State $EndAddr \gets (lower\_bound + upper\_bound) >> 1$
					\If{IsCachelineEvicted(array, Addr, EndAddr)}
						\State $upper\_bound \gets EndAddr - 1$
						\State $ConflictAddr \gets EndAddr$
					\Else
						\State $lower\_bound \gets EndAddr + 1$
					\EndIf
				\EndWhile
				\State $ConflictAddrList$.insert($ConflictAddr$)
			\EndFor
			\State \Return $ConflictAddrList$
			\EndFunction
		\end{algorithmic}
	\label{algo:find-cacheline-conflict}
\end{algorithm}

%\subsection{Summarizing VRAM Channel ID Permutation Pattern}

\begin{algorithm}[tb]
	\caption{Mark VRAM channel IDs in the VRAM space}
	\begin{algorithmic}\footnotesize
		\Function{MarkMemoryChannel}{Addr}
		\State $DramConflictAddrs \gets [], Cnt \gets 0$
		\For{$Addr'=Addr+1; Cnt < NeedNum; Addr'++$}
		\If{$IsDRAMBankConflicted(Addr, Addr')$}
		\State $DramConflictAddrs$.insert($Addr'$), $Cnt++$
		\EndIf
		\EndFor
		\State $CacheConflictAddrs \gets [], Cnt \gets 0$
		\For{$Addr' \in DramConflictAddrs$}
		\State $AddrList \gets FindCacheConflictAddrs(Addr')$
		\State $CacheConflictAddrs$.insert($AddrList$)
		\EndFor
		\State $MarkedAddressList \gets []$
		\For{$Addr' \in VRAM\ Address\ Space$}
		\State $tmp \gets array[Addr']$
		\For{$MarkedAddr \in CacheConflictAddrs$}
		\State $tmp \gets array[MarkedAddr]$
		\EndFor
		\State $t = TimeElapse(tmp \gets array[Addr'])$
		\If{$t > THRESHOLD$} \Comment{L2 cache miss occurs}
		\State $MarkedAddressList$.insert($Addr'$)
		\EndIf
		\EndFor
		\State \Return $MarkedAddressList$
		\EndFunction
	\end{algorithmic}
	\label{algo:mark-memory-channel}
\end{algorithm}

\subsection{Findings of VRAM Channel Reverse Engineering} \label{appendix:reverse-engineering-findings}

\begin{figure}[t]
	\centering
	%\vspace{-.15in}
	\subfloat[GTX 1080]
	{
		\includegraphics[width=0.3\columnwidth]{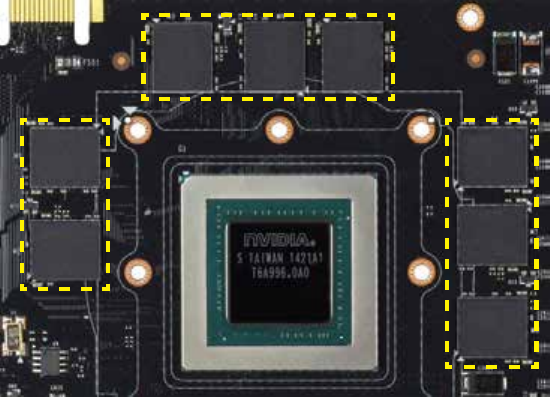}
		\label{fig:todo}
	}
	\subfloat[Tesla P40]
	{
		\includegraphics[width=0.3\columnwidth]{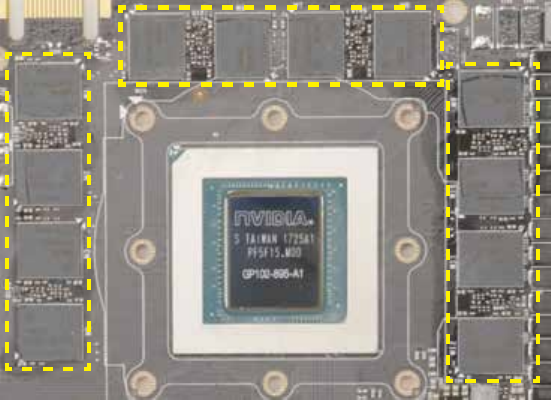}
		\label{fig:todo}
	}
	\subfloat[RTX A2000]
	{
		\includegraphics[width=0.3\columnwidth]{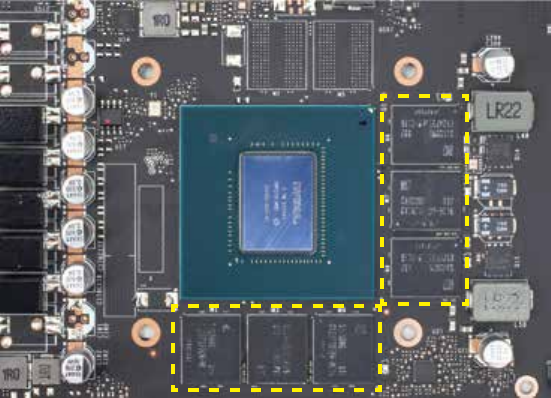}
		\label{fig:todo}
	}
	\caption{Disassembling GPUs: (a) GTX 1080 (8 VRAM channels), (b) Tesla P40 (12 VRAM channels), and (c) RTX A2000 (6 VRAM channels). The number of GDDR chips (dashed yellow rectangles) in each GPU is equal to the number of VRAM channels.}
\label{fig:gpu-disassembly-photo}
\end{figure}

\begin{figure}[tb]
	\centering
	\includegraphics[width=\columnwidth]{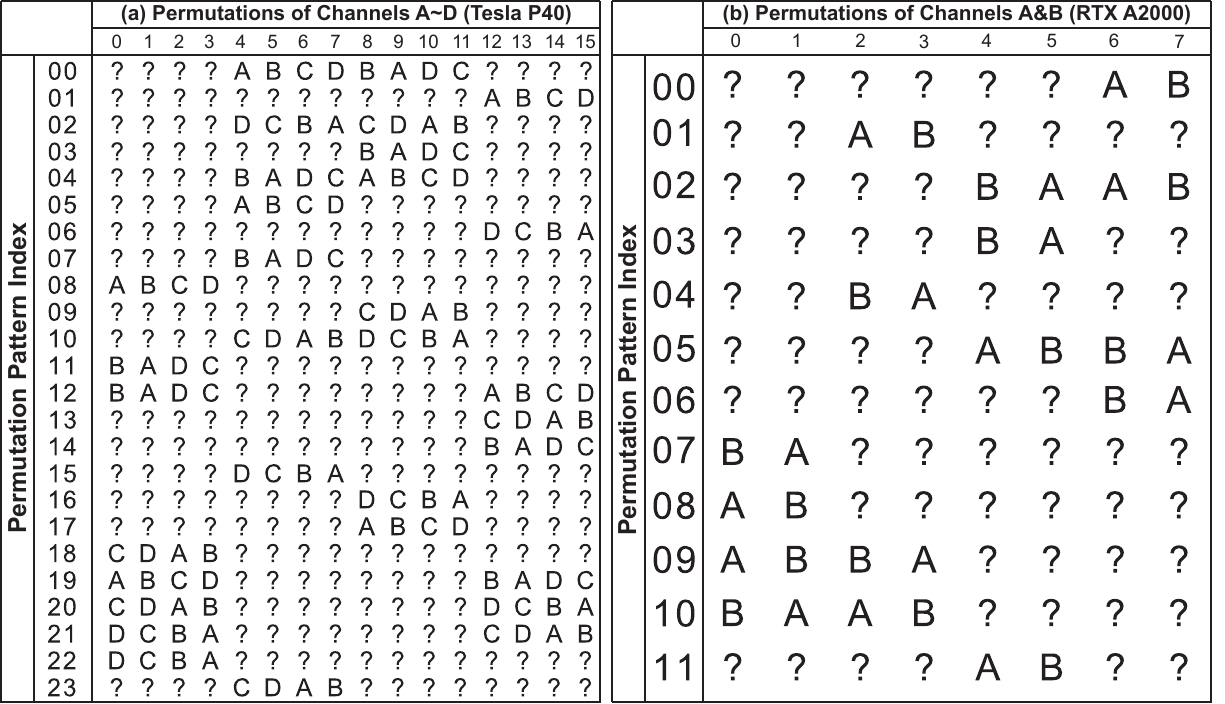}
	\caption{VRAM channel permutations of Tesla P40 (channels \texttt{A} $\sim$ \texttt{D}) and RTX A2000 (channels \texttt{A} $\sim$ \texttt{B}). \textbf{Note:} \texttt{?} denotes VRAM channels not in \texttt{A} $\sim$ \texttt{D} (\texttt{A} $\sim$ \texttt{B}).}
	\label{fig:full-reverse-engineering-results}
\end{figure}

Fig.~\ref{fig:gpu-disassembly-photo} presents the GDDR memory unit layouts of GTX 1080, Tesla P40, and RTX A2000, which can cross-validate the \# of VRAM channels of the three GPUs. We present the full reverse engineering results of VRAM channel permutations of Tesla P40 and RTX A2000 in Fig.~\ref{fig:full-reverse-engineering-results}.

\subsection{Rules of Deciding the Coloring Granularity}\label{appendix:decide-coloring-granularity}
Here, we conclude the rules deciding the maximum coloring granularity for VRAM channel isolation. The minimum and maximum coloring granularities, the maximum number of contiguous VRAM channels of each GPU are listed in Tab.~\ref{tab:granularity-min-max}. We have the following principles: 1) \textit{Minimum coloring granularity} = \textit{Channel partition size}; and 2) \textit{Maximum coloring granularity} = (Max \# contiguous VRAM channels) KiB.

If we allocate $2^N$ ($N$ = 0,1,2,...) channels to a task, the coloring granularity should be min($2^N$, \textit{Maximum coloring granularity}) KiB. If we want to allocate $N$ (not a power of 2) channels to any tasks, the granularity can only be 1 KiB.

\begin{table}[tb]
	\centering
	\scriptsize
	\begin{tabular}{c|cccc}
		\hline
		\textbf{GPU}        & \makecell[c]{\textbf{Minimum}\\\textbf{Coloring}\\\textbf{Granularity}} & \makecell[c]{\textbf{Maximum}\\\textbf{Coloring}\\\textbf{Granularity}} & \makecell[c]{\textbf{\# Contiguous}\\\textbf{VRAM}\\\textbf{Channels}} & \makecell[c]{\textbf{\# VRAM}\\\textbf{Channels}} \\ \hline
		GTX 1080   & 1 KiB                          & 4 KiB                          & 4                               & 8                \\
		Tesla P40  & 1 KiB                          & 4 KiB                          & 4                               & 12               \\
		%Tesla V100 & 1 KiB                          & 8 KiB                          & 8                               & 32               \\
		RTX A2000  & 1 KiB                          & 2 KiB                          & 2                               & 6                \\ \hline
		%RTX A5500  & 1 KiB                          & 2 KiB                          & 2                               & 12                \\ \hline
	\end{tabular}
	\caption{Minimum and maximum coloring granularities, the maximum number of contiguous VRAM channels, and the number of VRAM channels of three GPUs.}
\label{tab:granularity-min-max}
\end{table}


\clearpage
\bibliographystyle{ACM-Reference-Format}
\bibliography{references}

\end{document}